\documentstyle[prl,aps,epsf,twocolumn,draft]{revtex}

\begin{document}

\title{Relation between Raman spectra and Structure of Amorphous Silicon}

\author{R.L.C. Vink}
\address{Instituut Fysische Informatica,
         Utrecht University, Princetonplein 5, 3584 CC Utrecht,
         the Netherlands \protect\cite{vinkadd}}

\author{G.T. Barkema}
\address{Theoretical Physics,
         Utrecht University, Princetonplein 5, 3584 CC Utrecht,
         the Netherlands }

\author{W.F. van der Weg}
\address{Debye Institute,
         Utrecht University, P.O. Box 80,000, 3508 TA Utrecht,
         the Netherlands }

\date{\today}

\maketitle

\begin{abstract} In 1985, Beeman, Tsu and Thorpe established an almost linear
relation between the Raman transverse-optic (TO) peak width $\Gamma$ and the spread
in mean bond angle $\Delta\theta$ in {\it a}-Si. This relation is often used to
estimate the latter quantity in experiments. In the last decade, there has been
significant progress in the computer generation of sample networks of amorphous
silicon. Exploiting this progress, this manuscript presents a more accurate
determination of the relation between $\Gamma$ and $\Delta\theta$ using 1000-atom
configurations. Also investigated and quantified are the relations between the TO
peak frequency and the ratio of the intensities of the transverse-acoustic (TA) and
TO peak, both as functions of $\Delta\theta$. As $\Delta\theta$ decreases, the
TA/TO intensity ratio decreases and the TO peak frequency increases. These
relations offer additional ways to obtain structural information on {\it a}-Si from
Raman measurements. \end{abstract}

\section{Introduction}

Many structural properties of {\it a}-Si, such as defect concentration and
variation in mean bond angle, are difficult to determine experimentally.
This is because it is impossible to measure directly the coordinates of the
atoms in {\it a}-Si. However, important information on the structure of
{\it a}-Si can be obtained indirectly, through a number of experimental
techniques. These techniques include neutron, x-ray and Raman
scattering, electron-spin resonance, and x-ray photo-absorption.
Compared to other methods, Raman scattering is more sensitive to small
changes in the short-range order of {\it a}-Si. For this reason, Raman
measurements on {\it a}-Si are frequently used to obtain structural
information~\cite{maley,smith,bermejo,ivanda}.

The experimental Raman spectra of {\it a}-Si show two distinct peaks, at about
150 cm$^{-1}$ and 480 cm$^{-1}$, associated with the transverse acoustic (TA) and
the transverse optic (TO) vibrational modes, respectively. Certain features in
the Raman spectrum are highly sensitive to the structural properties of the {\it
a}-Si sample. For example, the width of the TO peak is related to the
root-mean-square bond-angle variation $\Delta\theta$ in the amorphous
network\cite{sinke}.

In several computational studies\cite{beeman,tsu,wong}, the relation
between $\Gamma$ and $\Delta\theta$ was quantified. All studies
indicate a broadening of the TO peak with increasing $\Delta\theta$,
but there is no quantitative agreement. Beeman's linear relation,
$\Gamma=15+6\Delta\theta$, which dates back to 1985, is often used by
experimentalists to determine $\Delta\theta$ from Raman measurements.
Here, $\Gamma$ is in cm$^{-1}$ and $\Delta\theta$ in degrees.

Beeman derived his relation using nine structural models of {\it a}-Si.  Of these
models, five were generated from the same 238-atom, hand-built model by Connell
and Temkin\cite{connell}, which contains only even-membered rings. In contrast,
all simulations on {\it a}-Si find an abundance of five- and seven-fold rings.
Moreover, these five Connell-Temkin models are statistically dependent and not
periodic, consequently containing a large fraction of surface atoms. Experimental
values of $\Delta\theta$, based on the radial distribution function of {\it a}-Si
obtained in neutron-diffraction studies, range from 9.9 to 11.0
degrees\cite{fortner}.  Of the nine structural models used by Beeman, only three
exhibit values of $\Delta\theta$ in this range.  New techniques to generate {\it
a}-Si structures, such as ART~\cite{art1,art2,art3}, as well as more powerful
computers, have made it possible to generate larger and more realistic {\it a}-Si
systems via computer simulation.

Also the description of the Raman scattering process has improved.
Beeman used the bond polarizibility model proposed by Alben {\it et
al.}\cite{alben}, which dates back to 1975.  Characteristic for this
model is the inclusion of three weighting parameters, whose values must
be set somewhat arbitrarily.  Several studies have indicated that the
values originally proposed by Alben yield an incorrect value for the
depolarization ratio~\cite{bermejo,ivanda}. These studies therefore
propose different weights.  Since then, other polarizibility models
have been proposed, for example by Marinov and Zotov~\cite{marinov1}.

In this manuscript, we will re-investigate the relation between $\Gamma$ and
$\Delta\theta$ by computer simulation. This simulation is based on a large number
of 1000-atom, periodic configurations, with structural properties (radial
distribution function, spread in mean bond angle) that are in excellent agreement
with experiment. Furthermore, recent advances in neutron scattering techniques
have made it possible to directly compare the bond polarizibility models to
experiment~\cite{lannin}. We therefore also include a detailed comparison of the
model of Alben and the model of Marinov and Zotov to experiment.

Additionally, we present two other methods to obtain structural
information from the Raman spectrum. The TA/TO intensity
ratio~\cite{maley} and the location of the TO-peak~\cite{zotov1} are
believed to be directly related to $\Delta\theta$; these relations will
be quantified.

The outline of this paper is as follows. In section~\ref{sect:method}, we
explain the generation of the {\it a}-Si configurations used in this study. We
then discuss how the Raman spectrum is obtained from these configurations. The
results and conclusions are presented in sections \ref{sect:results} and
\ref{sect:conclusions}, respectively.

\section{Method}
\label{sect:method}

To calculate Raman spectra, three ingredients are required:  (1) a
potential describing the atomic interactions in the sample, (2) a
continuous random network representing a realistic sample of {\it
a}-Si, and (3) a model assigning Raman activities to the vibrational
eigenmodes of the sample.

In the present work, we use a modified version of the Stillinger-Weber
potential for all calculations. This potential has the same functional
form as the original SW potential~\cite{stillinger}, but with different
parameters. The parameters were chosen specifically to describe {\it
a}-Si, see Ref.~\cite{vink}.

\subsection{Sample generation}

To study the effect of $\Delta\theta$ on the Raman spectrum, we
require a number of {\it a}-Si configurations with varying values of
$\Delta\theta$. These configurations were generated using the
activation-relaxation technique (ART)~\cite{art1,art2,art3}. As was
shown in previous studies~\cite{art1,art2}, ART yields structures in
good agreement with experiment. They display a low density of
coordination defects, a narrow bond-angle distribution and an
excellent overlap with the experimental radial distribution function
(RDF). The method is outlined below:

\begin{enumerate}

\item Initially, 1000 atoms are placed at random in a periodic cubic cell; the
configuration is then relaxed at zero pressure.

\item The configuration is annealed using ART. One ART move consists of two
steps: (1) the sample is brought from a local energy minimum to a nearby
saddle-point (activation), and (2) then relaxed to a new minimum with a local
energy minimization scheme including volume optimization, at zero pressure.  
The new minimum is accepted with a Metropolis probability at temperature
$T=0.25$ eV.

\item Every 50 ART moves, up to approximately five ART moves per atom when
the energy has reached a plateau, the configuration is stored.
For 1000-atom samples, this procedure yields a set containing 100 samples.
\end{enumerate}

This procedure is repeated nine times, generating nine statistically independent
sets of 100 correlated configurations each.  For each set, we found that
$\Delta\theta$ ranges from approximately 10$^\circ$ for the well-annealed
configurations, to approximately 14$^\circ$ for the poorly annealed
configurations.

\subsection{Calculation of Raman spectra}

We focus on the reduced Raman spectrum $I(\omega)$, with thermal and harmonic
oscillator factors removed. This spectrum is a function of frequency $\omega$
and of the form
\begin{equation}
  I(\omega) = C(\omega) g(\omega),
\end{equation}
where $g(\omega)$ is the vibrational density of states (VDOS) and
$C(\omega)$ a coupling parameter, which depends on frequency and on
the polarization (HH or HV) of the incident light used in the Raman
experiment.

To calculate the VDOS, the hessian is calculated. Diagonalization of
the hessian gives the frequencies of the vibrational modes, from
which the VDOS is obtained.

The function $C(\omega)$ is obtained from the polarizibility tensor
$\alpha(\omega)$. In terms of $\alpha(\omega)$, the coupling parameter
for HH and HV Raman scattering becomes $C_{HH}(\omega_p)=7 G^2+45A^2$
and $C_{HV}(\omega_p)=6G^2$, respectively, with $A$ and $G^2$ given by
the tensor invariants
\begin{equation}
  A = \frac{1}{3} \left[ \alpha_{11} + \alpha_{22} + \alpha_{33} \right]  \\
\end{equation}
and
\begin{eqnarray}
  G^2 &=& 3\left[ \alpha_{12}^2 + \alpha_{23}^2 + \alpha_{31}^2 \right] + \\
      &\frac{1}{2}& \left[ (\alpha_{11}-\alpha_{22})^2 + 
                           (\alpha_{22}-\alpha_{33})^2 +
                           (\alpha_{33}-\alpha_{11})^2 \right]; \nonumber
\end{eqnarray}
see for instance Ref.~\cite{bell}.

The form of the polarizibility tensor $\alpha(\omega)$ still needs to
be specified; this is the most uncertain part of the calculation.
Several models have been proposed, amongst which the commonly used
model of Alben {\it et al.}~\cite{alben} and the more recent model of
Marinov and Zotov~\cite{marinov1}.

In the model of Alben, a cylindrical symmetry of the individual bonds is
assumed and each bond is treated independently as a homopolar, diatomic
molecule. Three different forms for the bond polarizibility tensor are
introduced:
\begin{eqnarray}
  \alpha_1(\omega_p) &=& \sum_{l,\Delta} \vec{u}_l \cdot \vec{r}_\Delta
    \left[ \vec{r}_\Delta \vec{r}_\Delta - \frac{1}{3} {\bf I} \right], \\
  \alpha_2(\omega_p) &=& \sum_{l,\Delta} \vec{u}_l \cdot \vec{r}_\Delta
    \left[ \frac{1}{2} \left( \vec{r}_\Delta\vec{u}_l + \vec{u}_l\vec{r}_\Delta
    \right) - \frac{1}{3} {\bf I} \right], \\
  \alpha_3(\omega_p) &=& \sum_{l,\Delta} (\vec{u}_l \cdot \vec{r}_\Delta) {\bf I}. 
\end{eqnarray}
Here, the summation runs over all atoms $l$ in the sample and their nearest
neighbors $\Delta$, $\vec{r}_\Delta$ is the unit vector from the equilibrium
position of atom $l$ to the nearest neighbor $\Delta$, $\vec{u}_l$ is the
displacement vector of atom $l$ when it is vibrating in mode $p$ and $\bf I$
is the unit dyadic. The total polarizibility tensor $\alpha$ is a weighted
sum of the three terms, i.e. $\alpha = B_1 \alpha_1 + B_2 \alpha_2 + B_3
\alpha_3$. As was stated in the introduction, the precise choice of the
weights $B_i$ is somewhat arbitrary and this is the major shortcoming of the
model. Several studies have indicated that mechanisms 1 and 3 provide the
main contribution to the Raman scattering process; these propose to use
$B_1:B_2:B_3$ proportional to $2:0:1$, respectively~\cite{bermejo,ivanda}. In
this paper, we will use this set of weights.

The model of Marinov and Zotov (MZ) has no free parameters. In this
model, the bond polarizibility is expressed as a sum of three
components; two components parallel to the bond arising from bonding
and non-bonding electrons and a third component perpendicular to the
bond, see Ref~\cite{marinov1}. Under these assumptions, the
polarizibility tensor takes the form:
\begin{eqnarray}
  \alpha(\omega_p) &=& \sum_m r_m^3 \left[ 
    \left( \vec{b}_m \cdot \vec{r}_m \right) \vec{r}_m \vec{r}_m \right. \nonumber \\
    &&+ \frac{1}{2} \left. 
    \left( \vec{b}_m \vec{r}_m + \vec{r}_m \vec{b}_m \right) \right]. 
\end{eqnarray}
Here, the summation runs over all bonds $m$ in the sample, $\vec{r}_m$ is a
unit vector parallel to the bond, $r_m$ is the bond length and $\vec{b}_m$
is defined as $\vec{u}_j-\vec{u}_i$; where $\vec{u}_i$ and $\vec{u}_j$ are
the displacement vectors of atoms $i$ and $j$ constituting the $m$th bond,
when vibrating in mode $p$.

The coupling parameter for {\it a}-Si has also been determined
experimentally~\cite{lannin} through neutron scattering methods.
According to this experiment, the coupling parameter is a slowly
increasing function of frequency. We will use the experimental result
of Ref.~\cite{lannin} to test the validity of both the model of Alben
and the MZ model.

\section{Results}
\label{sect:results}

First, in section~\ref{sect:coupling}, we compare the polarizibility models of
Alben and Marinov and Zotov to experiment. In the subsequent sections, we
investigate the relation between the spread in the bond angle $\Delta\theta$ and:
\begin{enumerate}
  \item Raman TO peak width,
  \item Raman TO peak position, and 
  \item Raman TO/TA intensity ratio.
\end{enumerate}
We show results for HV polarized light only; this is the usual experimental
situation. Results for HH polarized light have also been obtained and are
available upon request.

\subsection{Raman coupling parameter}
\label{sect:coupling}

\begin{figure}
  \begin{center}
  \epsfxsize=8cm
  \epsfbox{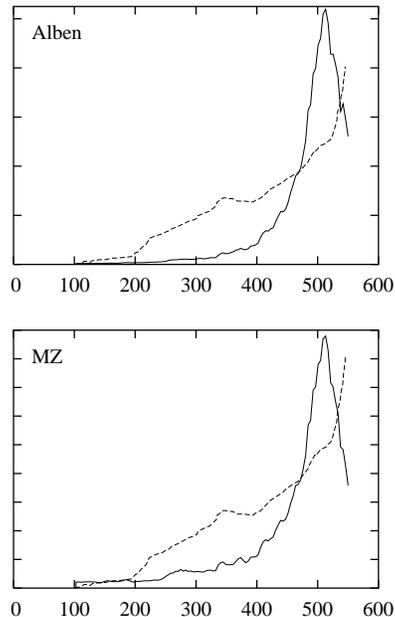}
  \end{center}
\caption{HV Raman Coupling parameter for {\it a}-Si calculated using the model of
Alben model (top) and the MZ model (bottom). The dashed line is the experimental
result taken from Ref.~\protect\cite{lannin}. Frequency is in cm$^{-1}$ and
all curves are area-normalized.}
\label{fig:coupling} 
\end{figure}

The solid curves in Fig.~\ref{fig:coupling} show the HV Raman coupling
parameter for {\it a}-Si calculated using the model of Alben (top) and the
MZ model (bottom) for the bond polarizibility. The experimental result of
Ref.~\cite{lannin} is also shown (dashed). For this calculation, we used a
well-annealed, 1000-atom configuration with $\Delta\theta=10.0^\circ$, since
this will most closely resemble the experimental sample. We have checked
that the general features of the curves in Fig.~\ref{fig:coupling} do not
depend on the details of the configuration used: a number of other,
well-annealed, configurations, with $\Delta\theta$ ranging from
10.0$^{\circ}$ to 11.0$^{\circ}$, gave similar results.

Fig.~\ref{fig:coupling} shows that both models yield an increasing coupling
parameter for frequencies up to around 500 cm$^{-1}$. This is in qualitative
agreement with experiment. For higher frequencies, the model calculations
predict a sharp decrease in the coupling parameter. This is not confirmed by
experiment.

The quantitative agreement with experiment is rather poor, especially in the
low-frequency regime; both models provide substantially less activity in this
regime than observed in experiment. For this reason, Raman spectra calculated
using either of the two models yield TA peak amplitudes far below experimental
values. This, in our opinion, is their major shortcoming.

\begin{figure}   
  \begin{center}
  \epsfxsize=8cm
  \epsfbox{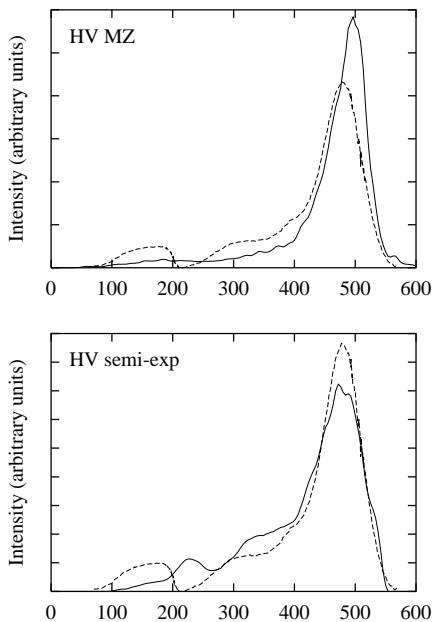}
  \end{center} 
\caption{ {\bf top:} Reduced HV Raman spectrum for {\it a}-Si calculated using the
MZ model (solid). {\bf bottom:} Reduced HV Raman spectrum calculated using the
semi-experimental approach; the VDOS is obtained by simulation, the coupling
parameter is taken from Ref.~\protect\cite{lannin}. The dashed line in both graphs
shows an experimental reduced Raman spectrum taken from
Ref.~\protect\cite{berntsen}. Frequency is in cm$^{-1}$ and all spectra are
area-normalized.}
\label{fig:spectra} 
\end{figure}

This point is further illustrated in the top graph of Fig.~\ref{fig:spectra}, where
we show the Raman spectrum calculated using the MZ model (solid) and an
experimental spectrum (dashed) taken from Ref.~\cite{berntsen}. The experimental
spectrum was obtained from ion-implanted {\it a}-Si which had been annealed at
500$^\circ$C for two hours. Agreement between model and experiment, particularly in
the low-frequency regime, is poor.

Given the overall poor performance of both the model of Alben and the MZ model, it
may be feasible to follow a semi-experimental approach in which a computer
generated {\it a}-Si sample is used to calculate the VDOS and experimental data is
used to describe the coupling parameter. This approach is justified because several
studies have indicated that changes in the Raman spectrum are due to changes in the
VDOS and not to changes in the coupling parameter~\cite{berntsen,maley2}. The
results of this approach are illustrated in the bottom graph of
Fig.~\ref{fig:spectra}. Here, we show the Raman spectrum obtained by multiplying a
computer-generated VDOS with experimental coupling parameter data (solid). To
calculate the VDOS, we used a well-annealed {\it a}-Si sample with
$\Delta\theta=10.0^\circ$; the experimental coupling parameter was taken from
Ref.~\cite{lannin}. The dashed line shows again the experimental Raman spectrum
taken from Ref.~\cite{berntsen}. Agreement with experiment has improved
substantially.

Of the two models considered here, the MZ model provides slightly more activity in
the low-frequency regime than the model of Alben; comparison with experiment would
therefore favor the MZ model. However, given the overall poor performance of both
models, we will also show in the following sections results obtained using the
semi-experimental approach.

\subsection{Raman TO peak width vs. $\Delta\theta$}
\label{sect:pw}

The intensity of the Raman TO peak decreases abruptly on the high frequency side,
but not on the low frequency side. Beeman therefore defines $\Gamma$ as twice the
half-width at half the maximum height on the high frequency side of the TO peak,
as a meaningful parameter to specify the TO peak width~\cite{beeman}. In this
paper, we use the same definition.

\begin{figure}
  \begin{center}
  \epsfxsize=8cm
  \epsfbox{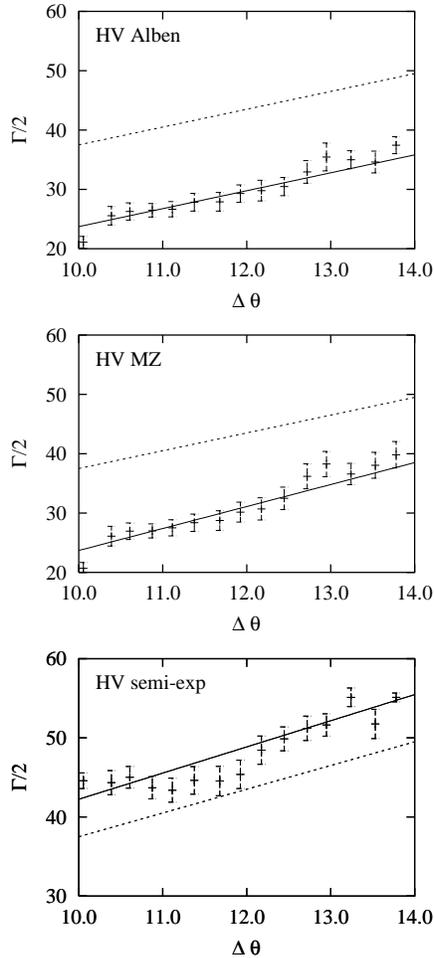}
  \end{center}
\caption{HV Raman TO peak width $\Gamma/2$ as a function of $\Delta\theta$, calculated
using the model of Alben (top), the MZ model (middle) and the semi-experimental approach
(bottom). The solid lines are linear least-squares fits; the dashed line is the result of
Beeman: $\Gamma/2=7.5+3\Delta\theta$. The units of $\Gamma/2$ and $\Delta\theta$ are
cm$^{-1}$ and degrees, respectively.}
\label{fig:pwHV}
\end{figure}

The solid lines in Fig.~\ref{fig:pwHV} show the relation between $\Gamma$ and
$\Delta\theta$ for HV polarized light, derived using the model of Alben (top), the MZ
model (middle) and the semi-experimental approach (bottom). Also shown is the result
obtained by Beeman (dashed).

Both the model of Alben and the MZ model produce similar results; linear least
square fits yield the equations $\Gamma/2=3.0\Delta\theta-6.5$ and
$\Gamma/2=3.7\Delta\theta-13.3$, respectively. Compared to the result of Beeman,
$\Gamma/2=3.0\Delta\theta+7.5$, we see agreement on the sensitivity (i.e.~slope of
the lines) of $\Gamma$ to $\Delta\theta$, but not on the overall offset
(i.e.~intercepts of the lines). The same holds for the result obtained in the
semi-experimental approach; least square fitting yields
$\Gamma/2=3.3\Delta\theta+9.2$ in that case.

\subsection{Raman TO peak position vs. $\Delta\theta$}

\begin{figure}
  \begin{center}
  \epsfxsize=8cm
  \epsfbox{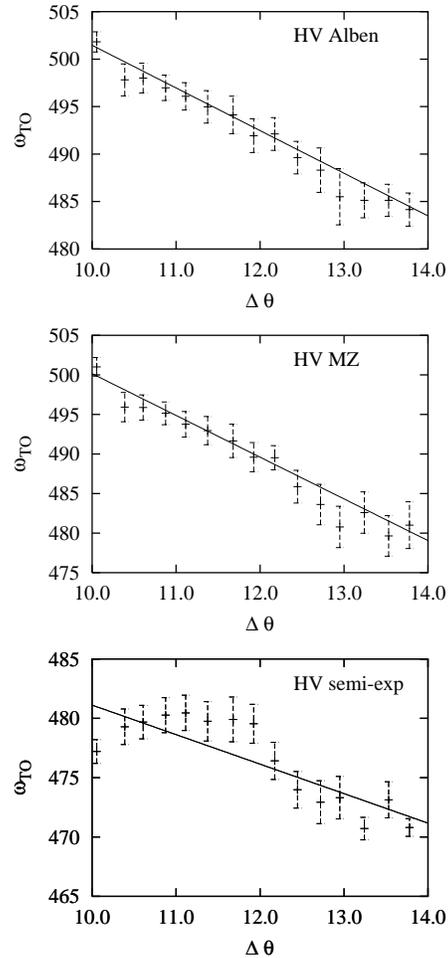}
  \end{center}
\caption{Reduced HV Raman TO peak position as a function of $\Delta\theta$ for {\it
a}-Si, calculated using the model of Alben (top), the MZ model (middle) and the
semi-experimental approach (bottom). The peak position and $\Delta\theta$ are given
in cm$^{-1}$ and degrees, respectively. The solid lines are least-squares fits
through the datapoints.}
\label{fig:to}
\end{figure}

As another way to obtain structural information on {\it a}-Si from its Raman
spectrum, we investigate the relation between the TO peak frequency ($\omega_{TO}$)
and $\Delta\theta$. Fig.~\ref{fig:to} shows the relation between $\omega_{TO}$ and
$\Delta\theta$ for HV polarized light, derived using the model of Alben (top), the
MZ model (middle) and the semi-experimental approach (bottom). 

According to Fig.~\ref{fig:to}, $\omega_{TO}$ shifts to higher frequency as
$\Delta\theta$ decreases. However, agreement with experiment for both the model of
Alben and the MZ model is poor. Experimental Raman spectra of well-annealed {\it
a}-Si samples yield $\omega_{TO}$ in the order of 480 cm$^{-1}$~\cite{berntsen}.
For well-annealed configurations, for which $\Delta\theta$ ranges from 9.9 to 11.0
degrees, the models exceed the experimental value by around 20 cm$^{-1}$.

The semi-experimental approach is in much better agreement with experiment; a
linear fit yields the equation $\omega_{TO}=-2.5\Delta\theta+505.5$ which for
$\Delta\theta=10.0^\circ$ leads to $\omega_{TO}=480.5$ cm$^{-1}$.

The cause of this is that in the model of Alben and the MZ model,
$\omega_{TO}$ is determined by the peak in the coupling parameter,
whereas in the semi-experimental approach, it is determined by the
VDOS.

\subsection{Raman TA/TO intensity ratio vs. $\Delta\theta$}
\label{sect:ta}

\begin{figure}
  \begin{center}
  \epsfxsize=8cm
  \epsfbox{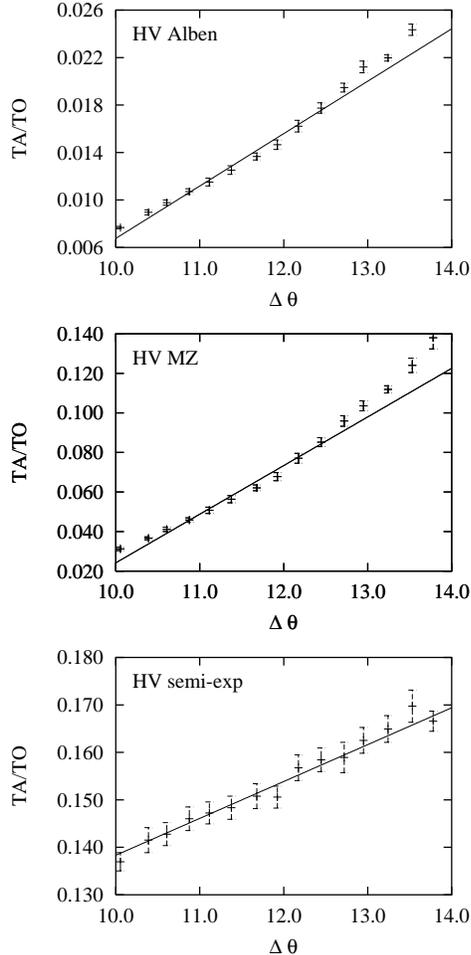}
  \end{center}
\caption{HV Raman TA/TO intensity ratio as a function of $\Delta\theta$ for {\it
a}-Si, calculated using the model of Alben (top), the MZ model (middle) and the
semi-experimental approach (bottom). The solid lines are least-squares fits.
$\Delta\theta$ is in degrees.}
\label{fig:ta}
\end{figure}

Next, we confirm that the TA/TO intensity ratio is directly related to $\Delta\theta$.
Fig.~\ref{fig:ta} shows the relation between reduced Raman TA/TO intensity and
$\Delta\theta$ for HV polarized light for the model of Alben (top), the MZ model (middle)
and the semi-experimental approach (bottom)

From Fig.~\ref{fig:ta}, we see that the TA/TO intensity ratio increases with increasing
$\Delta\theta$. The increase is approximately linear.

HV Raman experiments on well-annealed {\it a}-Si samples yield a reduced TA/TO
intensity ratio around 0.11~\cite{berntsen}. For both the model of Alben and the MZ
model, the HV TA/TO ratio is of order $10^{-2}$, i.e.~one order of magnitude below
the experimental value. This is consistent with the earlier finding that both
models underestimate the Raman activity in the low-frequency regime of the
spectrum, see section~\ref{sect:coupling}.

The semi-experimental approach, on the other hand, yields a TA/TO ratio of 0.14
for $\Delta\theta=10.0^\circ$, in better agreement with experiment.

\section{Conclusions}
\label{sect:conclusions}

We have generated nine independent sets of 1000-atom samples of {\it a}-Si that
display a variety of short-range order; the spread in nearest-neighbor bond
angles ranges from 10 to 14 degrees. For these samples, the HV Raman spectra are
calculated.  To describe the Raman scattering process, we have used the earlier
bond polarizibility model of Alben {\it et al.}, the more recent model of Marinov
and Zotov as well as experimental data taken from Ref.~\cite{lannin}.

Comparison to experiment shows that both the model of Alben and the
MZ model greatly underestimate the Raman activity in the
low-frequency regime of the spectrum. This makes these models less
suitable to describe low-frequency features of the Raman spectrum,
for instance the TA peak. Of the two models considered here, the MZ
model is closer to experiment. However, for a more accurate
calculation of Raman spectra, we propose a semi-experimental
approach. In this approach, the VDOS is obtained in computer
simulation and experimental data is used to describe the coupling
parameter.

As ways to obtain structural information on {\it a}-Si from its Raman
spectrum, we have investigated the relation between the TO peak-width
$\Gamma$ and $\Delta\theta$, as well as the relations between the TO peak
frequency and the TA/TO intensity ratio as functions of $\Delta\theta$.

According to our results, where we used the semi-experimental approach, $\Gamma$
and $\Delta\theta$ are related by $\Gamma/2=3.3\Delta\theta+9.2$ for HV polarized
light.  Here, $\Gamma$ is in cm$^{-1}$ and $\Delta\theta$ in degrees. Comparing
this to the result of Beeman ($\Gamma/2=3\Delta\theta+7.5$), we find that our
result is similar.

Our results also show a shift of the Raman TO peak frequency ($\omega_{TO}$)
towards higher frequency, as $\Delta\theta$ decreases. In the semi-experimental
approach, a linear least-square fit yields $\omega_{TO}=-2.5\Delta\theta+505.5$
for HV polarized light. Here, $\omega_{TO}$ is in cm$^{-1}$ and $\Delta\theta$ in
degrees. According to this equation, the shift of $\omega_{TO}$ is approximately
7.5 cm$^{-1}$, going from unannealed {\it a}-Si $(\Delta\theta \approx 13^\circ)$
to annealed {\it a}-Si $(\Delta\theta \approx 10^\circ)$. This is in quantitative
agreement with experiment~\cite{berntsen}.

Finally, we have shown that the reduced Raman TA/TO intensity ratio $(I)$ is
directly related to $\Delta\theta$; $I$ decreases linearly with decreasing
$\Delta\theta$. Using the semi-experimental approach, we obtain the relation
$I=0.0078\Delta\theta+0.0606$, where $\Delta\theta$ is in degrees.

\end{document}